\documentstyle[twocolumn,aps]{revtex}

\begin{document}
\draft

\title{Wigner crystallization in the two electron quantum dot}
\author{A.~Matulis}
\address{Institute of Semiconductor Physics, Go\v{s}tauto 11,
2600 Vilnius, Lithuania}
\author{F.~M.~Peeters\cite{AddFMP}}
\address{Departement Natuurkunde, Universiteit Antwerpen (UIA),
Universiteitsplein 1, B-2610 Antwerp, Belgium
}
\date{\today}
\maketitle

\begin{abstract}
Wigner crystallization can be induced in a quantum dot by
increasing the effective electron-electron interaction through
a decrease of the electron density or by the
application of a strong magnetic field. We show that the ground
state in both cases is very similar but the energy scales are very
different and therefore also the dynamics.
\end{abstract}

\pacs{PACS numbers: 73.20.Dx, 85.30.Vw, 03.65.-w}%

\section{Introduction}

During the last ten years quantum dots have attracted a lot of interest
both experimentally and theoretically \cite{johnson95}. Recently,
much attention was paid to the investigation of quantum dots in strong
perpendicular magnetic fields where a rich structure of cusps
and steps in the chemical potential $\mu(N,B)$
(as a function of confined electron number $N$ and applied magnetic
field strength $B$) was observed \cite{ooster99}. The above structure
is related to changes in the ground state electron density in
the presence of a magnetic field, and it initiated numerous theoretical
investigations after Chamon and Wen \cite{wen94} proposed the
quantum dot edge reconstruction. The electron density in a
quantum dot is the result of the interplay of the repulsive
character of the electron interaction and the attractive forces
caused by the confining potential, the magnetic field and the
electron exchange interaction. If the magnetic field is strong
enough, the overlap between the electron wave functions becomes less,
the electron interaction will dominate, and a ring of electrons at the
dot edge is formed. It was also found that if the magnetic field is increased
further the above electron ring becomes unstable, and a ground state
with a broken rotational symmetry appears. The possibility of the appearance
of spin waves~\cite{reim97} and charge density waves
\cite{wing99,reim99} have also been reported.

A quantum dot with a density profile consisting of
rings with lumps reminds us to the
Wigner crystal \cite{mull96,renn99} which is the ground state
of the classical electron system in a dot \cite{bed94}
in the absence of a magnetic field. In the latter  case the Wigner
crystal occurs when the potential energy (the inter-electron
interaction and the confinement potential) dominates over the
electron kinetic energy. This is just the classical limit in a quantum
problem. Therefore, classical or quasi-classical
methods should be adequate for the description of the Wigner crystal.
But from a first sight it appears that such a classical limit
is not reached when a strong magnetic field is applied. The above
mentioned electron density reconstruction was revealed when the
magnetic field is so strong that the electrons occupy the lowest
Landau level. Thus, the electron motion quantization is essential,
and consequently, the kinetic energy exceeds the potential energy
due to the Coulomb interaction.
Nevertheless, due to the degeneracy of the Landau level this large
electron kinetic energy is actually frozen out, and Wigner
crystallization results from the same potential energy as
in the classical case without magnetic field.
Actually the above crystallization is a result of different
energy scales in the electron system under consideration.

The purpose of the present paper is, by using an exactly solvable
model of two electrons in a dot, to illustrate the conditions
under which a Wigner crystal can be formed in the case of a strong
magnetic field ($B$), and to compare this quantum Wigner crystal
with the classical zero magnetic field one. The paper is organized as
follows. In Sect. II we present our model. Sect. III gives an introduction
to the quasi-classical approach for the $B=0$ case. Sect. IV discusses
the general $B\ne 0$ case. Our conclusions are presented in Sect. V.

\section{Model}

We consider two electrons with effective masses $m^*$ moving in the
$z=0$ plane and which are confined by a two-dimensional harmonic
potential of characteristic frequency $\omega_0$. A magnetic field is
applied in the $z$ direction and described by the vector potential
chosen in the symmetric gauge ${\bf A}=[{\bf B}\times {\bf r}]/2$.
The corresponding Hamiltonian can be separated into two parts
\begin{equation}
H = H_R + H_r,
\end{equation}
(see, for instance \cite{merkt91,mat94}) which
represents the center-of-mass and relative motion (with corresponding
coordinates ${\bf R} = ({\bf r}_1+{\bf r}_2)/2$ and
${\bf r}={\bf r}_1-{\bf r}_2$). In dimensionless form
those parts can be written as follows:
\begin{mathletters}
\begin{eqnarray}
H_R &=& -\frac{1}{4}\nabla_R^2 + \left\{1+\frac{B^2}{4}
\right\}R^2 -\frac{iB}{2}[{\bf R}\times{\nabla_R}]_z, \\
H_r &=& - \nabla_r + \frac{1}{4}\left\{1+\frac{B^2}{4}\right\}r^2
-\frac{iB}{2}[{\bf r}\times\nabla_r]_z + \frac{\lambda}{r}.
\end{eqnarray}
\end{mathletters}
The system energy is measured in $\hbar\omega_0$ units, and
the coordinates in $a_0=\sqrt{\hbar/m^*\omega_0}$ units.
The symbol $\lambda = a_0/a_B$ stands for the electron interaction
coupling constant which is the ratio of the characteristic
dot dimension $a_0$ and the Bohr radius $a_B=\epsilon\hbar^2/m^*e^2$.
The magnetic field strength $B$ is measured in $\Phi_0/\pi a_0^2$
units where $\Phi_0=\pi\hbar c/e$ is the magnetic flux quantum.

We will concentrate ourselves to the study of the coordinate wave
function part of the ground singlet state
\begin{equation}\label{ftot}
\Psi({\bf r}_1,{\bf r}_2) = \Phi({\bf R})
\psi({\bf r}),
\end{equation}
the corresponding electron density
\begin{eqnarray}\label{dens}
\rho({\bf r}) &=& \int d^2r_1\int d^2r_2
|\Psi({\bf r}_1,{\bf r}_2)|^2
\hat{\rho}({\bf r}) \nonumber \\
&=& \int d^2r_1\int d^2r_2|\Psi({\bf r}_1,{\bf r}_2)|^2
\sum_{n=1}^2\delta({\bf r}-{\bf r}_n) \nonumber \\
&=& 2\int d^2r_1 |\Phi({\bf r}+{\bf r}_1/2)|^2
|\psi({\bf r}_1)|^2
\end{eqnarray}
and the correlation function
\begin{eqnarray}\label{corr}
K({\bf r},{\bf r}') &=& \int d^2r_1 \int d^2r_2
|\Psi({\bf r}_1,{\bf r}_2)|^2 \nonumber \\
&&\phantom{m}\times\left\{\hat{\rho}({\bf r})\hat{\rho}({\bf r}')-
\delta({\bf r}-{\bf r}')\hat{\rho}({\bf r})\right\}
\nonumber \\
&=& 2|\Phi(\{{\bf r}+{\bf r}'\}/2)|^2
|\psi({\bf r}-{\bf r}')|^2.
\end{eqnarray}
It is the latter function which enables one to determine whether, or not
the system is in the Wigner crystal state.

\section{Zero magnetic field case}

We consider first the simplest case in which no magnetic field
is applied. Then the center-of-mass motion part is trivial. It just
represents the two dimensional harmonic oscillator motion which has
no relation to the Wigner crystallization problem. Its ground
state energy is $E_R=1$ with the corresponding wave function
\begin{equation}\label{fcmm}
\Phi({\bf R}) = \sqrt{\frac{2}{\pi}}\exp(-R^2).
\end{equation}
Due to the cylinder symmetry the relative motion wave function
part can be written as
\begin{equation}\label{frel}
\psi({\bf r}) = \frac{1}{\sqrt{2\pi}}\exp(im\varphi)R(r),
\end{equation}
where the radial function $R(r)$ is obtained by solving the one
dimensional eigenvalue problem as determined by the Hamiltonian
\begin{mathletters}\label{hamrel}
\begin{eqnarray}
H_r = -\frac{1}{r}\frac{d}{dr}r\frac{d}{dr} + V(r,\lambda), \\
\label{vrel}
V(r,\lambda) = \frac{m^2}{r^2} + \frac{1}{4}r^2
+ \frac{\lambda}{r}.
\end{eqnarray}
\end{mathletters}

As was already mentioned strong electron correlation
(and the Wigner crystal as well) occurs when the potential
energy dominates over the electron kinetic energy,
i.~e.~when $\lambda\to\infty$. In this interesting limit
the eigenvalue problem is strongly simplified and can be solved
by analytical means. Indeed, in the case of $\lambda\to\infty$
the potential (\ref{vrel}) has a minimum close to the point
$r_0=(2\lambda)^{1/3}$ (see solid curve in Fig.\ \ref{fig1}).
The potential can be expanded into a $(r-r_0)$ series (dashed curve in
Fig.\ \ref{fig1})
\begin{equation}\label{pp}
V(r,\lambda) \approx \frac{3}{4}(2\lambda)^{2/3} +
\frac{3}{4}(r-r_0)^2 + \frac{m^2}{(2\lambda)^{2/3}},
\end{equation}
which actually coincides with the quasi-classical expansion into
a $\lambda^{-2/3}$ series \cite{mat94}.
The form of the above expansion clearly indicates the
energy scales of the different excitations in the quantum dot.
The first term ($\lambda^{2/3}$) is just the classical dot
energy which can also be obtained using
the hydrodynamic approximation~\cite{zar94}.
In the $\lambda\to\infty$ limit this energy dominates. Thus, we are in
the region of the classical Wigner crystal. The solution of the
Schr\"{o}dinger equation with the harmonic term in the expansion
of the potential (\ref{pp}) leads to $\lambda$-independent equidistant
electron ring vibration excitations (shown in Fig.\ \ref{fig1} by
the horizontal dashed lines). Note that the separation of those
vibration levels are of order $1$, and consequently, much less
than the classical potential energy.
The last term in the expansion (\ref{pp}) describes
the rotation energy. It leads to a small (of order
$\sim\lambda^{-2/3}$) splitting of the rotation levels as is
shown by the encircled part in Fig.\ \ref{fig1} which is enlarged.
The spectrum is thus similar to the one of molecules with bands of
rotation levels, attached to each vibration level.

For the expanded potential (\ref{pp}) we obtain the following
quasi-classical radial ground state wave function,
\begin{equation}\label{fquasi}
P(r) = N \exp\left\{-a(r-r_0)^2\right\},
\end{equation}
where $a=\sqrt{3}/4$ and the normalization
$N=(2a/\pi)^{1/4}r_0^{-1/2}$.

Inserting the above relative motion wave function (\ref{fquasi})
together with the center-of-mass motion function (\ref{fcmm}) into
expressions (\ref{dens},\ref{corr}) we obtain the electron density
\begin{eqnarray}\label{dens0}
&&\rho(r,\varphi) = \frac{2N^2}{\pi^2} \int_0^{\infty}dr'r'
\exp(-2r^2-{r'}^2/2) \nonumber \\
&& \times \exp\left\{-2a(r'-r_0)^2\right\} \int_0^{2\pi} d\varphi'
\exp\left\{2rr'\cos(\varphi-\varphi')\right\} \nonumber \\
&=& \frac{4N^2}{\pi} \int_0^{\infty}dr'r'
\exp\left\{-2r^2-{r'}^2/2 -2a(r'-r_0)^2\right\} \nonumber \\
&& \times I_0(2rr')
\approx \frac{N^2}{\pi}\sqrt{\frac{2\gamma}{a}}
\exp\left\{-\gamma(r-r_0/2)^2\right\},
\end{eqnarray}
and the correlation function
\begin{eqnarray}
K(\mathbf{r},\mathbf{r}') &=& \frac{4N^2}{\pi}
\exp\left\{-(\mathbf{r}+\mathbf{r}')^2/2\right\}
\nonumber \\ && \phantom{mmm} \times
\exp\left\{-2a(|\mathbf{r}-\mathbf{r}'|-r_0)^2\right\}.
\end{eqnarray}
In expression (\ref{dens0}), $I_0$ is the Bessel function
and $\gamma=8a/(1+4a) \approx 1.268$.

A schematic picture of the above functions is shown in
Fig.\ \ref{fig2} by the shadowed regions. In Fig.~\ref{fig2}(a)
the electron density is depicted which is mainly concentrated
on a thin ring. Notice that it has no lumps, and exhibits the
cylindrical symmetry of the system Hamiltonian.
Wigner crystal state can be seen in the correlation function
\cite{maksym93,yan99} which is plotted in Fig.~\ref{fig2}(b).
The latter is actually the same as the conditional probability
distribution. One electron is fixed in position $\mathbf{r}'$ (indicated
by a cross in Fig.~\ref{fig2}(b)) and the density of the other electron is
then mainly concentrated in the opposite position which clearly indicates
the presence of Wigner crystallization in this quasi-classical
limit.

\section{Strong magnetic field case}

Due to the symmetric gauge which preserves
the cylindrical symmetry, both (i.~e.~center-of-mass
and relative) wave functions can be written as a product of the orbital
exponent and the radial part as in expression (\ref{frel}).
The corresponding radial Hamiltonians (for center-of-mass and relative motions)
can be presented as follows:
\begin{mathletters}\label{magbase}
\begin{eqnarray}
H_R &=& -\frac{1}{4}\frac{d}{dR}R\frac{d}{dR} +
\frac{1}{4}\left(\frac{M}{R}-BR\right)^2 + R^2, \\
H_r &=& -\frac{1}{r}\frac{d}{dr}r\frac{d}{dr} +
\left(\frac{m}{r}-\frac{Br}{4}\right)^2 + \frac{1}{4}r^2
+ \frac{\lambda}{r}.
\end{eqnarray}
\end{mathletters}
Because we are interested in the asymptotic limit
$B\to\infty$, it is convenient to scale the variables
and the Hamiltonian as follows
\begin{equation}\label{magscale}
{\bf r} \to {\bf r}B^{-1/2}, \quad
{\bf R} \to {\bf R}B^{-1/2}, \quad
H\to HB,
\end{equation}
in order to have the expansion parameter $B^{-1}$ explicitly in our
problem. Inserting (\ref{magscale}) into expression (\ref{magbase})
we arrive at the following Hamiltonian:
\begin{mathletters}
\begin{eqnarray}
H_R &=& -\frac{1}{4}\frac{d}{dR}R\frac{d}{dR} +
\frac{1}{4}\left(\frac{M}{R}-R\right)^2\!\!\!\!
+V_R(R,B), \\
H_r &=& -\frac{1}{r}\frac{d}{dr}r\frac{d}{dr} +
\left(\frac{m}{r}-\frac{r}{4}\right)^2
+V_r(r,B),
\end{eqnarray}
\end{mathletters}
where in the asymptotic region $B\to\infty$ the terms
\begin{mathletters}
\begin{eqnarray}
V_R(R,B) &=& \frac{R^2}{B^2}, \\
\label{magpotrel}
V_r(r,B) &=& \frac{r^2}{4B^2} + \frac{\lambda}{r\sqrt{B}}
\end{eqnarray}
\end{mathletters}
can be treated as small perturbations.

The solution of the Schr\"{o}dinger equations for the zero
order Hamiltonians (without the terms $V_R$ and $V_r$) leads to
the degenerate Landau levels where the ground state has energy
$E_R=E_r=1/2$ labeled by integer positive momentum ($M$
and $m$) values. Mathematically this degeneracy is a consequence
of the equivalence of the zero center-of-mass Hamiltonian potential
$(M/R-R)^2/4$ which is shown in Fig.\ \ref{fig3}(a) where
the orbital momentum is indicated by the corresponding integers.
For any momentum $M$ the potential curve has a minimum
equal to zero at the position $R_{\mathrm{min}}=\sqrt{M}$.
A similar potential is obtained for the zero order relative motion
equation, and will therefore not be discussed.

Next we take the perturbation terms into account.
For the case of center-of-mass motion the potential is shown in
Fig.~\ref{fig3}(b). The potential is composed of the same curves
as in the case without perturbation, but they are now moved
up by the amount $R^2/B^2$ shown in Fig.~\ref{fig3}(b) by the dotted
curve. The lowest minimum is obtained for $M=0$ which implies
that the center-of-mass motion wave function is the same
as in the case without magnetic field (\ref{fcmm}).

This is not the case for the relative motion where
all potential curves are shifted up by the amount $V_r(r,B)$
as shown in Fig.~\ref{fig3}(c) by the dotted curve. According
to Eq.~(\ref{magpotrel}) this curve increases for both $r\to 0$
and $r\to\infty$ and reaches a minimum value at
\begin{equation}\label{equ2}
r_{\mathrm{min}}= (2\lambda)^{1/3}\sqrt{B}.
\end{equation}
Minimizing the potentials with respect to the relative angular
momentum for $r=r_{\mathrm{min}}$ we obtain the corresponding
orbital momentum
\begin{equation}
m = \left(\frac{\lambda}{4}\right)^{2/3}B.
\end{equation}
Consequently, the relative motion wave function is peaked
on the ring of radius $r_0$. When the magnetic field
strength increases the radius $r_0$ tends to infinity,
while the orbital number is growing as well. Expanding
the potential with the lowest minimum in the vicinity of the
equilibrium point (\ref{equ2}) we find
\begin{equation}
\left(\frac{m}{r}-\frac{r}{4}\right)^2
\approx \frac{(r-r_{\mathrm{min}})^2}{4},
\end{equation}
which means that the thickness of the ring remains constant.

The layout of the energy spectrum in the large magnetic field limit
is shown in Fig.~\ref{fig4}.
Notice that the spectrum has two different energy scales as
in the $B=0$ case. But the physical meaning of those scales is
quite different.
The largest energy scale is the electron kinetic energy.
And the rotation levels are split due to the interplay of the Coulomb
interaction between the electrons and the confinement potential.

Going back to our original units before the scaling (\ref{magscale}) we
find that the ring radius is $r_{\mathrm{min}}=(2\lambda)^{1/3}$,
which is identical to the zero magnetic field case.
But now the thickness of the ring is $\sim B^{-1/2}$, and
it tends to zero as the magnetic field strength approaches
infinity.

Surprisingly, we arrived at the same situation as was found for
the case without magnetic field. The relative motion wave function
is located on a ring whose diameter greatly exceeds its
thickness. Thus the conclusions of previous section concerning
Wigner crystallization in the quantum dot are also valid in
the strong magnetic field case. While for $B=0$ the Wigner
crystal state is reached for $\lambda \gg 1$ we find that
a sufficiently strong magnetic field can crystallize the system
for any $\lambda\ne 0$.

\section{Conclusions}

The occurrence of Wigner crystallization depends on the ratio of the
distance between the electrons ($l$) and the characteristic
dimension of the single electron wave packet ($a$).
Actually this ratio is a measure of the electron density
in the quantum dot. In our case $l$ is given by the radius
$r_{\mathrm{min}}$ of the ring in the correlation function,
and $a$ coincides with its thickness. This ratio is
\begin{equation}
\chi = \frac{l}{a} = \frac{r_{\mathrm{min}}}{a}
= \frac{\lambda^{1/3}}{B^{-1/2}} = \lambda^{1/3}\sqrt{B},
\end{equation}
which in real units reads
\begin{equation}
  \chi = \left(\frac{a_0}{a_B}\right)^{1/3}
  \left(\frac{a_0}{l_B}\right).
\end{equation}
Here $l_B=\sqrt{\hbar c/eB}$ is the magnetic length.

The larger this ratio the more pronounced the Wigner
crystal state is.
Thus, both a strong magnetic field and strong interaction
favors Wigner crystallization, but in a different way.
The electron interaction makes the system less dense by enlarging
the inter-particle distance. The magnetic field also makes the
system effectively less dense but by compressing the single
electron wave packages.

Also we would like to note that mathematically in both cases
the static Wigner crystal properties can be calculated by
the same method, namely by the minimization of the classical potential
energy which is composed of the electron interaction and the confinement
potential. Nevertheless the physical meaning of that calculation
is quite different. In the strong electron interaction case the
potential energy dominates making the whole problem quasi-classical,
while in the case of strong magnetic fields, the problem
is essentially quantum mechanical (the Landau level energy is
dominating). But due to the degeneracy of the problem the system
is guided by the same potential energy as for the $B=0$ case
but with different energy scales.
Therefore, although the Wigner crystal configuration
is the same, one can expect different dynamics in the two limiting
cases.

\section*{Acknowledgments}

This work is supported by the Flemish Science Foundation,
IUAP-VI and the "Bijzonder Onderzoeksfonds van de Universiteit
Antwerpen".
One of us (F.~M.~P.) is a research director with FWO-Vl.
We acknowledge discussions with B.~Partoens, S.~Reimann and
D.~Pfannkuche.

\newpage

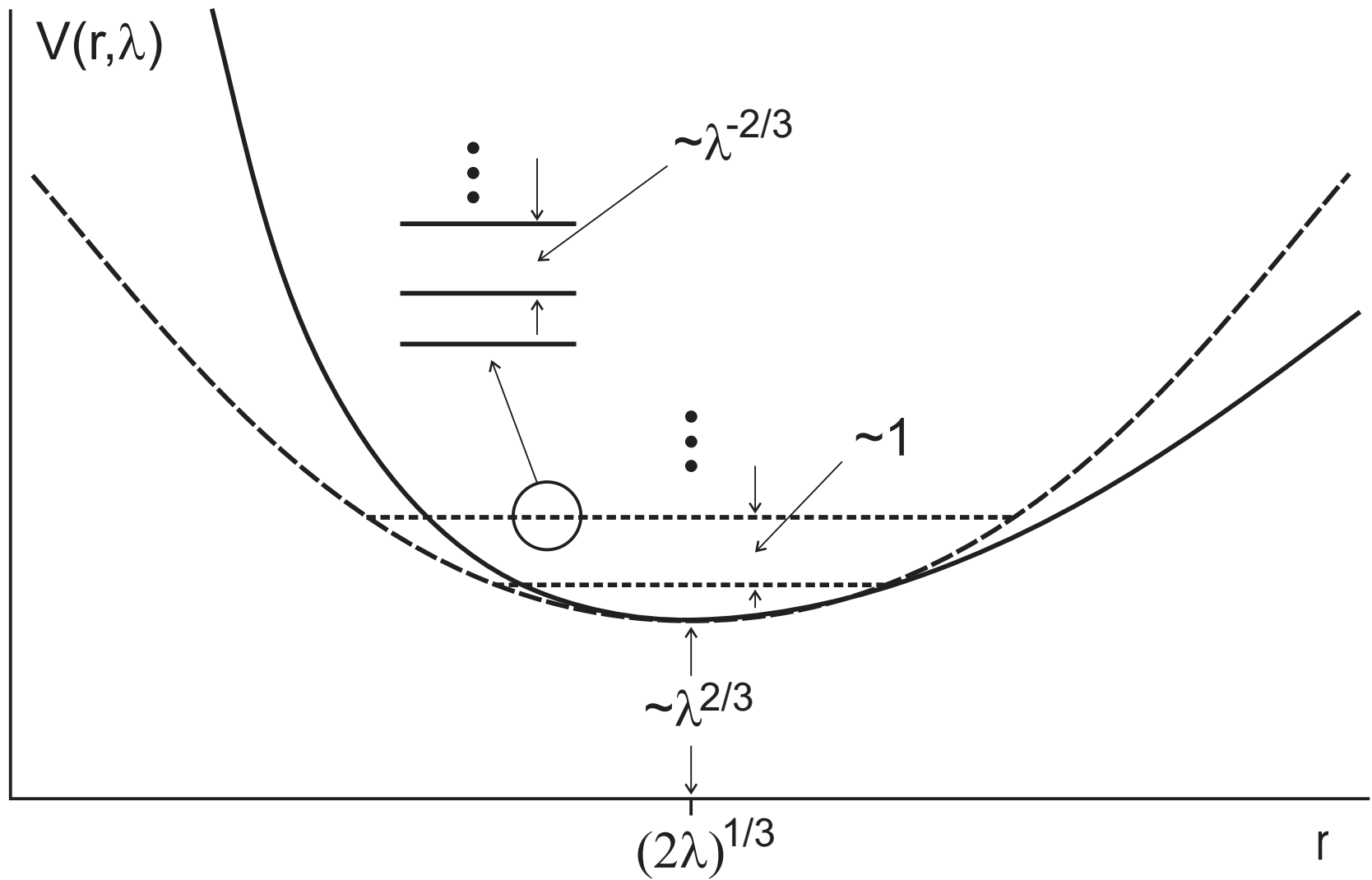
\begin{figure}
\caption{The potential energy for the relative motion (solid curve)
and its quadratic expansion (dashed curve) around its minimum (\ref{pp}).
The horizontal lines are the different vibration levels in units
of $\hbar\omega_0$.}
\label{fig1}
\end{figure}
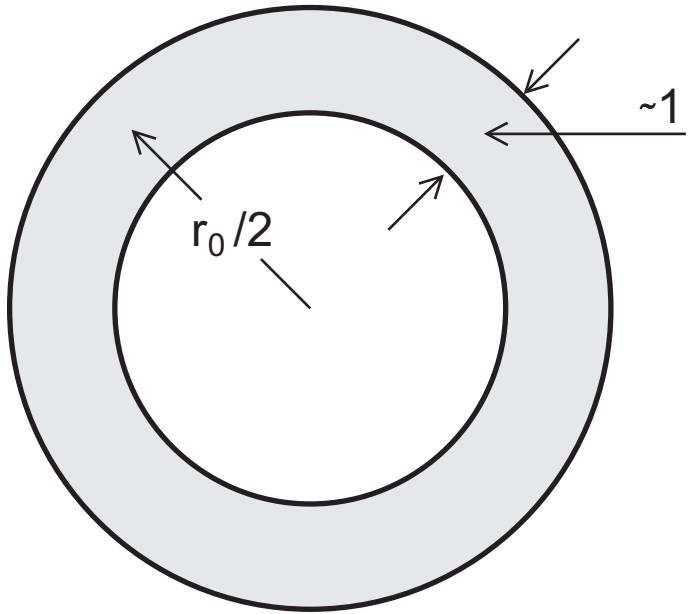
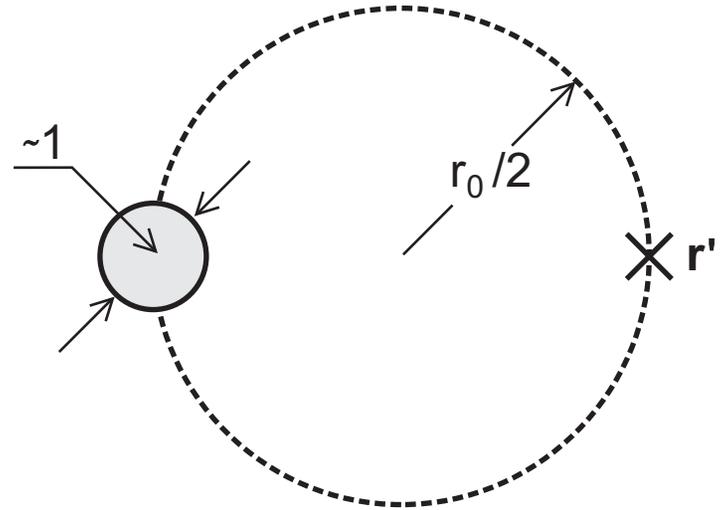
\begin{figure}
\caption{Schematic view of the electron density (a) and the electron
correlation function (b) in the quasi-classical limit.}
\label{fig2}
\end{figure}
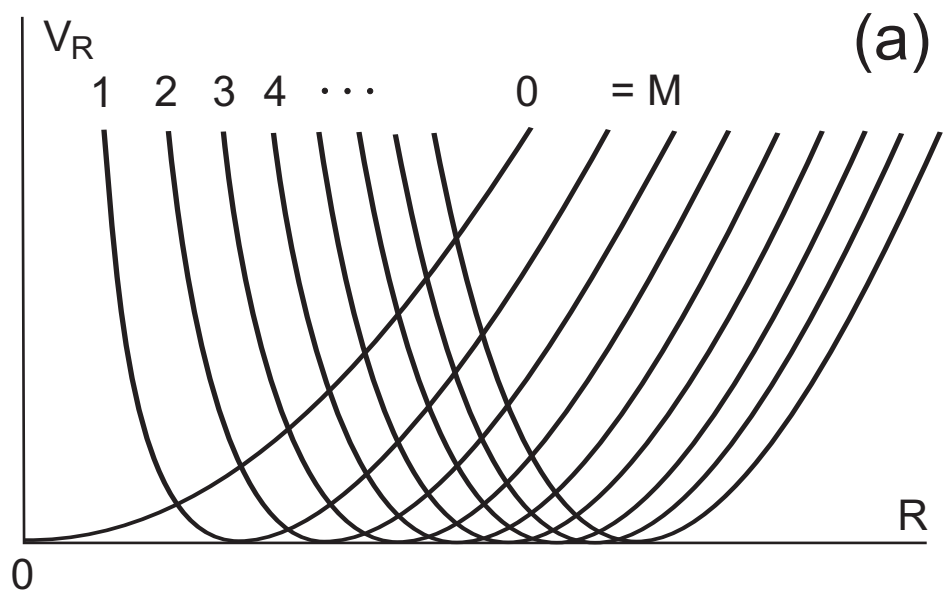
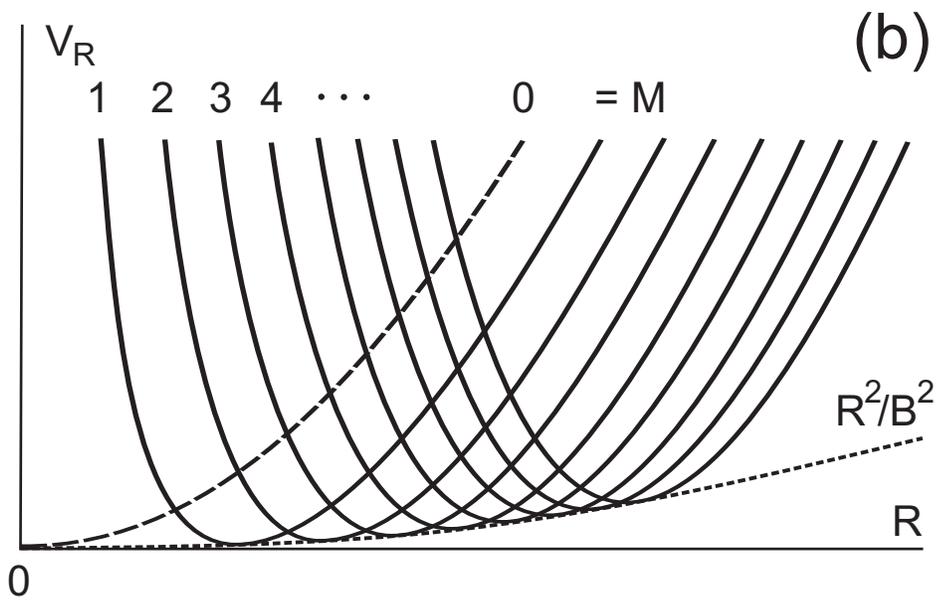
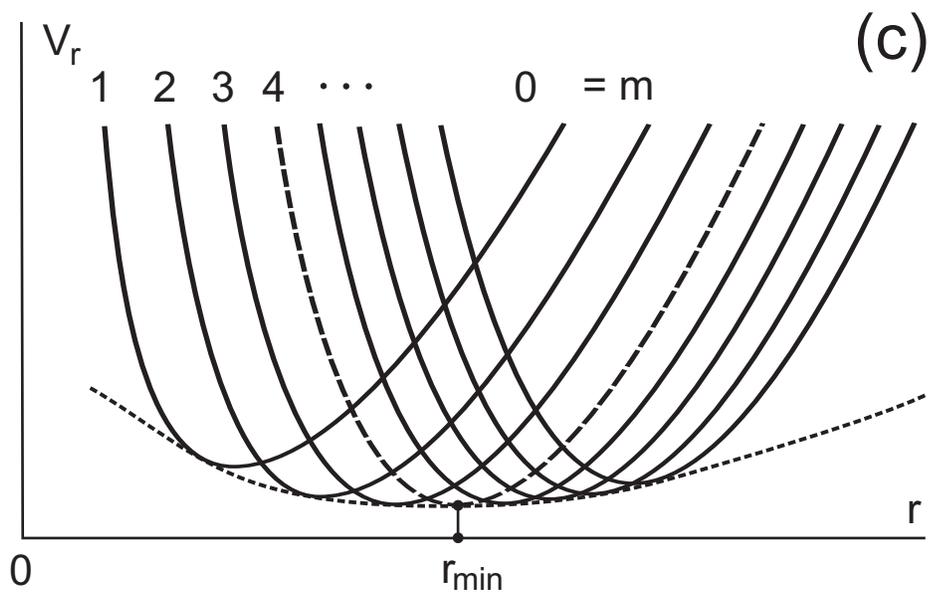
\begin{figure}
\caption{The potential energy curves for different angular momentum
in case of: (a) the non-interacting problem,  (b) the center-of-mass
motion including confinement, and (c) the relative motion including
confinement and electron-electron interaction.}
\label{fig3}
\end{figure}
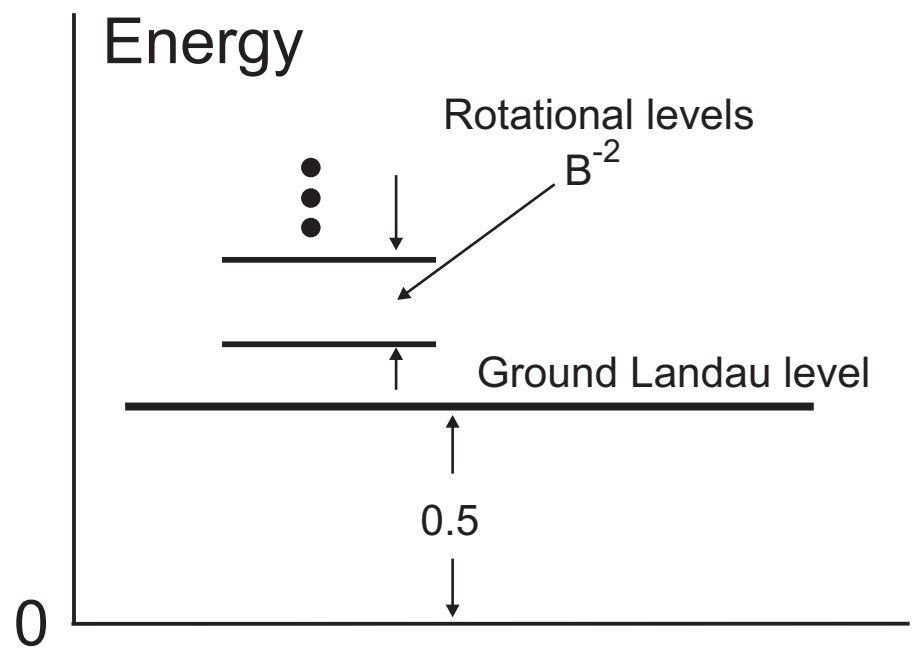
\begin{figure}
\caption{Energy spectrum in the limit of a strong magnetic field.
The energy is given in units of $\hbar\omega_c$.}
\label{fig4}
\end{figure}

\end{document}